\documentclass[aps,pre,twocolumn,amsmath,amssymb,superscriptaddress]{revtex4-1}

\usepackage{preamble}
\graphicspath{/Users/fergus/Dropbox/Fergus/writing/Figures}
\usepackage[normalem]{ulem}

\definecolor{aqua}{rgb}{0,0.588,1.0}
\definecolor{strawberry}{rgb}{1.0,0.0,0.5}

\definecolor{myblue}{rgb}{0.003921569,0.070588235,0.474509804}

\newcommand{\pecl}{\operatorname{\mathit{P\kern-.08em e}}}

\begin{document}

\title{Crystallisation and Polymorph Selection in Active Brownian Particles} 
\date{\today}

\author{Fergus J. Moore}
\email{\text{fergus.moore@bristol.ac.uk}}
\affiliation{Bristol Centre for Functional Nanomaterials, University of Bristol, Bristol BS8 1FD, United Kingdom}
\affiliation{H.H. Wills Physics Laboratory, Tyndall Ave., Bristol, BS8 1TL, UK}

\author{C. Patrick Royall}
\email{\text{paddy.royall@espci.fr}}
\address{Gulliver UMR CNRS 7083, ESPCI Paris, Universit\' e PSL, 75005 Paris, France.}
\affiliation{School of Chemistry, Cantock's Close, University of Bristol, BS8 1TS, UK}
\affiliation{H.H. Wills Physics Laboratory, Tyndall Ave., Bristol, BS8 1TL, UK}

\author{Tanniemola B. Liverpool}
\affiliation{School of Mathematics, University of Bristol, Bristol BS8 1UG, United Kingdom}

\author{John Russo}
\email{\text{john.russo@uniroma1.it}}
\affiliation{Department of Physics, Sapienza University of Rome, P.le Aldo Moro 5, 00185 Rome, Italy}
\affiliation{School of Mathematics, University of Bristol, Bristol BS8 1UG, United Kingdom}

\begin{abstract}
We explore crystallisation and polymorph selection in active Brownian particles with numerical simulation. In agreement with previous work  [Wysocki \emph{et al.} \emph{Europhys. Lett.}, \textbf{105} 48004 (2014)], 
we find that crystallisation is suppressed by activity and occurs at higher densities with increasing P\'{e}clet number ($\pecl$). While the nucleation rate decreases with increasing activity, the crystal growth rate increases due to the accelerated dynamics in the melt. As a result of this competition we observe the transition from a nucleation and growth regime at high $\pecl$ to 
 ``spinodal nucleation'' at low $\pecl$.
Unlike the case of passive hard spheres, where preference for FCC over HCP polymorphs is weak, activity causes the annealing of HCP stacking faults, thus strongly favouring the FCC symmetry at high $\pecl$.  When freezing occurs more slowly, in the nucleation and growth regime, this tendency is much reduced and we see a trend towards the passive case of little preference for either polymorph.
\end{abstract}

\maketitle

\section{Introduction}
\label{sectionIntroduction}

The field of Active Matter may be said to consider systems 
of organisms or artificial bodies that consume energy for self-propulsion \cite{ramaswamy2017}. On mesoscopic length scales ($nm$ to $\mu m$), it is concerned with describing 
the dynamics of biological microswimmers \cite{elgeti2015} such as bacteria and motile cells \cite{henkes2020}. The dynamics of active matter in unbounded, homogeneous, and low-Reynolds number environments are well described by active Brownian motion and run and tumble dynamics \cite{bechinger2016}. Observation of matter behaving according to this description has led to the discovery of unique dynamical phenomena such as motility-induced phase separation (MIPS)\cite{cates2015}, where bodies packed at densities greater than a critical volume fraction and with sufficient propulsion will separate into a dense phase and a dilute phase in the absence of attraction.\cite{marchetti2016}

Key to the development of a better theoretical understanding is to use simple models of active particles which capture some of the complex behaviour observed experimentally, for example collective motion and demixing \cite{vicsek1995,gregoire2004,wang2011,mognetti2013,redner2013,stenhammar2014,fodor2016,cates2015,pohl2014,zottl2014,tjhung2018}. In this context, simple model systems, such as active colloids, play an important role and these may be modelled with the use of active Brownian particles (ABPs) \cite{marchetti2016}.

Besides being the source of novel phenomena, activity can also fundamentally alter the nature of behaviour already observed in passive systems, such as crystallisation. Although certain aspects of crystallisation in passive colloidal systems, such as the nucleation rate at low supersaturation are still poorly understood \cite{auer2004,palberg2014,wood2018,schilling2010,radu2014,tateno2019,kawasaki2010,filion2011,russo2013,espinosa2019,fiorucci2020}, at higher supersaturation, where crystallisation occurs on the timescales accessible to brute force computer simulations, very good agreement is found between experiment and simulation \cite{taffs2013}. At higher colloidal volume fraction still, the barrier to nucleation falls so much that rather than conventional nucleation--and--growth, the system undergoes ``spinodal nucleation'', where, relative to the intrinsic structural relaxation time $\tau_\alpha$,  the timescale for crystallisation falls dramatically such that it is well below the relaxation time ~\cite{taffs2013,zaccarelli2009,valeriani2012,sanz2014,yanagishima2017}.

Another property of crystallising systems is polymorphism, i.e. the ability of a material to nucleate different crystalline phases, and whose understanding is fundamental to predict the structure of the growing nuclei. So far, our understanding of polymorphism is based on equilibrium thermodynamic principles, such as the Ostwald step rule of phases \cite{ostwald1897}, stating that the first solid formed is not the thermodynamically most stable, but the state nearest in free energy to the original state. For hard spheres, which may be said to constitute the passive equivalent of the ABP system that we consider,
two different crystalline polytypes are observed during nucleation: either face-centered-cubic (FCC) or hexagonal-close-packed (HCP), and the difference in all thermodynamic relevant quantities (such as free-energy, nucleation barrier, and stacking-free energy) between the competing polymorph are negligibly small (within $10^{-3}\,k_BT$ per particle for all cases)~\cite{woodcock1997,pronk1999}. Thermodynamics thus dictate that the early stages of nucleation should produce an almost equal amount of FCC and HCP for hard-spheres. We will show that activity can significantly alter this result. Studying the effect of activity is thus an important step towards understanding polymorphism in out-of-equilibrium situations.

The effect of activity upon crystallisation has been studied in the context of the effect on the state diagram \cite{mallory2017}. In both  
two \cite{bialke2012,digregorio2018} and three dimensions \cite{wysocki2014,stenhammar2014,omar2021,turci2021} the freezing line is found to move to higher area or volume fraction as a function of activity. In this sense, activity may be said to suppress crystallisation. The effect of activity on the process of nucleation has been studied via classical nucleation theory, in which a renormalised surface tension was found to provide reasonable agreement with simulation \cite{redner2016}. At higher activity in dimension $d=3$, the active fluid that coexists with a low-density active fluid through MIPS has a very high volume fraction \cite{omar2021,turci2021} and crystal nucleation requires rare fluctuations that exhibit the nearly close-packed volume fraction of the solid \cite{omar2021}. One intriguing and unexpected effect of activity upon crystallisation was the observation of annealing of grain boundaries in the case of the addition of a small quantity of active particles to an otherwise passive system \cite{vandermeer2016}.

To date, there have been relatively few experiments with active colloids at high density where crystallisation due to excluded volume interactions is seen \cite{pusey1986}. This is due in no small part to the difficulties in stabilising active colloids at high density against aggregation. However, recently this has begun to change and excluded volume interactions have driven ordering in a few experiments in two dimensions \cite{mauleonamieva2020,klongvessa2019,palacci2013,vanderlinden2019}. The study of 3d active colloids is in its infancy, however one system that has emerged of active multi--polar colloids \cite{sakai2020} does exhibit crystallisation to a variety of polymorphs also exhibited by related passive dipolar colloids \cite{yethiraj2003,colla2018}.

In this work we consider crystallisation regimes in a system of
active Brownian particles 
in three dimensions. In particular we investigate analogous behaviour to the
nucleation--and--growth and spinodal regimes observed in passive colloidal systems. Furthermore we find an unexpected polymorph selection phenomenon that is uniquely distinct from those observed in passive systems.

This article is organised as follows. In section~\ref{sectionMethods} we describe the methodology used for the simulation runs and the analysis of topological clusters in the fluid. Results are presented in section~\ref{sectionResults}, with subsections dedicated to the state diagram (\ref{sectionStateDiagram}), the dynamical and structural properties of the active fluid (\ref{sectionDynamicalStructuralResponse}), and the nucleation and crystal growth behaviour (\ref{sectionCrystalGrowthActivity}). We summarise our findings in section~\ref{sectionConclusion}.

\section{Methods}
\label{sectionMethods}

\subsection{Computer Simulations}
\label{sectionComputerSimulations}

We model active colloids as active Brownian particles, which propel with a constant velocity $V_{0}$, along their individual direction vectors $\boldsymbol{e}$, which in turn are subject to rotational diffusion. 
We implement this model through molecular dynamics simulations using a customised version of the open source LAMMPS package \cite{plimpton1995}, which integrates the following
equations of motion:

\begin{equation}
    \dot{\mathbf{r}}=V_{0} \mathbf{e}+\beta D_{t}\mathbf{F} +\sqrt{2D_{t}}\boldsymbol{\eta}
    \label{eqEom1}
\end{equation}

\begin{equation}
    \dot{\mathbf{e}}=\sqrt{2D_{r}}\boldsymbol{\xi} \times \mathbf{e}
    \label{eqEom2}
\end{equation}

\noindent
Here $\dot{\mathbf{r}}$ is the particle velocity, $V_{0}$ is the magnitude of the constant applied active velocity, and $\boldsymbol{F}$ is the inter-particle force. The thermal fluctuations promoting translational diffusion are included in the Gaussian white-noise term
$\boldsymbol{\eta}$, where $\langle\boldsymbol{\eta}\rangle=0$, and $D_{t}$ is the translational diffusion coefficient. Thermal noise driving rotational diffusion of the direction vector $\boldsymbol{e}$ is represented by $\boldsymbol{\xi}$,  where $\langle\boldsymbol{\xi}\rangle=0$, and $D_{r}$ is rotational diffusion coefficient. The two diffusion coefficients are related via $D_{t} = D_{r}\sigma^{2}/3$. For all simulations in this work $\beta = 5$, $m=1$, $\sigma=1$. 
Our measure of time is the characteristic 
rotational diffusion time $\tau_r = 1 / (2D_r)$ \cite{wysocki2014}.

The active particles are modelled as being similar to hard spheres and to achieve this we include a Weeks-Chander-Andersen (WCA) inter-particle potential in the force term in equation \eqref{eqEom1}, which takes the form:

\begin{equation}
\beta u_\mathrm{wca}(r) = \begin{cases}
4 \beta \varepsilon\left[\left(\frac{\sigma}{r}\right)^{12} - \left(\frac{\sigma}{r}\right)^{6}\right] + \varepsilon & r \leq 2^{\frac{1}{6}}\sigma \\
0 & r > 2^{\frac{1}{6}}\sigma   
\end{cases}
\label{eqWCA}
\end{equation}

\noindent
where $\varepsilon$ is the interaction energy, $r$ is the inter-particle distance.

Since we use the WCA interaction,
we cannot assume the hard particle diameter $\sigma$ to define a volume fraction. Furthermore, methods that determine an effective particle diameter such as Barker--Henderson effective hard sphere diameter \cite{barker1967}, may not hold outside of equilibrium systems. Therefore, as in ref.\cite{martin2021}, we use the total density $\rho = N / V$, where $N$ is the number of particles and $V$ is the volume of the system.  

We use the P\'{e}clet number to refer to the relative strength of the activity in the system, which we define as: $\pecl = V_{0}/\sigma D_{r}$. Throughout this work we keep $D_r$ constant at $D_r =1§$, and vary $\pecl$ by changing the propulsion velocity $V_0$. Previous studies \cite{stenhammar2014,martin2021} have shown that when the propulsion force from the activity increases relative to the repulsive force from the WCA interaction, the particles become softer. This change was observed to manifest in a shifting of the MIPS phase boundary to higher $\pecl$ and $\rho$, an effect that can be reduced by using an inter-particle potential that more closely resembles that of the hard sphere interaction \cite{martin2021}. However, as we focus only on crystallisation in this work, we perform simulations at low $\pecl$ and well below the MIPS phase boundary, and thus this effect in our WCA system is negligible.

To prepare the high density initial configurations for the simulations, we use the Lubachevsky--Stillinger algorithm \cite{lubachevsky1990}.
This takes a given box size and number of particles and slowly grows and displaces the particles from $\sigma=0.1$ until they reach $\sigma=1$ with minimal overlaps. To guard against the presence of any small but sufficient remaining particle overlaps, we perform a pre-run simulation with a soft potential:

\begin{equation}
u(r)=A\left[1+\cos \left(\frac{\pi r}{r_{c}}\right)\right] \quad r<r_{c}
\label{eqFatting}
\end{equation}

\noindent
where the constant A is ramped from 0 to 100 over and $r_c = 2^\frac{1}{6}$. This is run for $1.2 \tau_R$ without activity. Following this we perform our data collection runs with particles following the equations of motion outlined in (\ref{eqEom1}) and (\ref{eqEom2}), and the Weeks--Chandler--Anderson (WCA) inter-particle potential (\ref{eqWCA}). 

We perform all simulations in a periodic cubic box of dimension length $L=27.5 \sigma$, and vary the number of particles from 18000 to 24000 to explore a range of densities. 
This system size is such that the largest critical nucleus observed in this work was comprised of less than 4\% of the particles in the system, to avoid the finite size effects that have been studied for seeded nucleation in the NVT ensemble~\cite{rosales2020}.
For the determination of structure in longtime steady states we run for 7200$\tau_R$ and average over 10 independent configurations. For analysis of nucleation dynamics, we run for 600$\tau_r$ and average over 20 independent configurations.

\subsection{Dynamical Analysis}
\label{sectionDynamical}

The structural relaxation time $\tau_{\alpha}$ provides a useful metric through which we can understand the the effects of active systems on the verge of crystallisation. We compute $\tau_{\alpha}$ for various $\phi$ and $\pecl$, through calculation of the self part of intermediate scattering function:
\begin{equation}
  	F_{\mathrm{s}}(k, t)=\frac{1}{N}\left\langle\sum_{j=1}^{N} \exp \left[\mathrm{i} \vec{k} \cdot\left(\vec{r}_{j}(t)-\vec{r}_{j}(0)\right)\right]\right\rangle
\end{equation}
\noindent
where $\vec{k}$ is the wavevector $k=|\vec{k}|$, taken as $2\pi/\sigma$. We define $\tau_{\alpha}$ as $F_s(k,\tau_{\alpha}) = e^{-1}$.

\subsection{Topological Cluster Classification Analysis}
\label{sectionTopological}

Local structures identified in this work are identified by the Topological Cluster Classification (TCC) \cite{malins2013}. The TCC algorithm analyses structure through clusters. To identify a cluster the TCC uses a modified Voronoi construction to identify a bond network with a cutoff $r_c=1.8\sigma$ and a four-membered ring parameter $f_c=0.82$.  We identify clusters through calculation of the shortest path 3, 4, and 5 membered rings in the bond network. 
For non-crystalline clusters we consider only the minimum energy clusters of the Leonard-Jones interaction, specifically: 5A, 6A, 7A, 8B, 9B, 10B, 11C, 12B, and 13A \cite{taffs2010}. Here the numbers denote the number of particles in each cluster and the lettering signifies the cluster geometry \cite{doye1995}.
Furthermore, we use the TCC to identify crystal structure, where 13 particle FCC or HCP clusters are determined through a central particle and its 12 nearest neighbours.
We quantify the degree to which a particular structure appears in a configuration as the cluster population $N_c/N$, where $N_c$ is the number of particles in a given cluster, and $N$ is the number of particles in the system. It is important to note that a particle can belong to more than once cluster. For example, under certain conditions a particle can belong to both an FCC cluster and an HCP cluster, and when comparing such cluster populations, $N_c/N$ will not sum to 1.

\section{Results}
\label{sectionResults}

In this section we study the dynamical features of ABP 
at high density, 
where an ordered crystalline phase is found to spontaneously form in simulations. We will trace the boundaries of the crystal region, distinguishing between state points that nucleate through a \emph{nucleation and growth} mechanism, and those that display \emph{spinodal nucleation}. Through the Topological Cluster Classification (TCC) method we will distinguish between the FCC and HCP structures and consider the effects of activity on polymorph selection.

\begin{figure}
\includegraphics[width=0.8\linewidth]{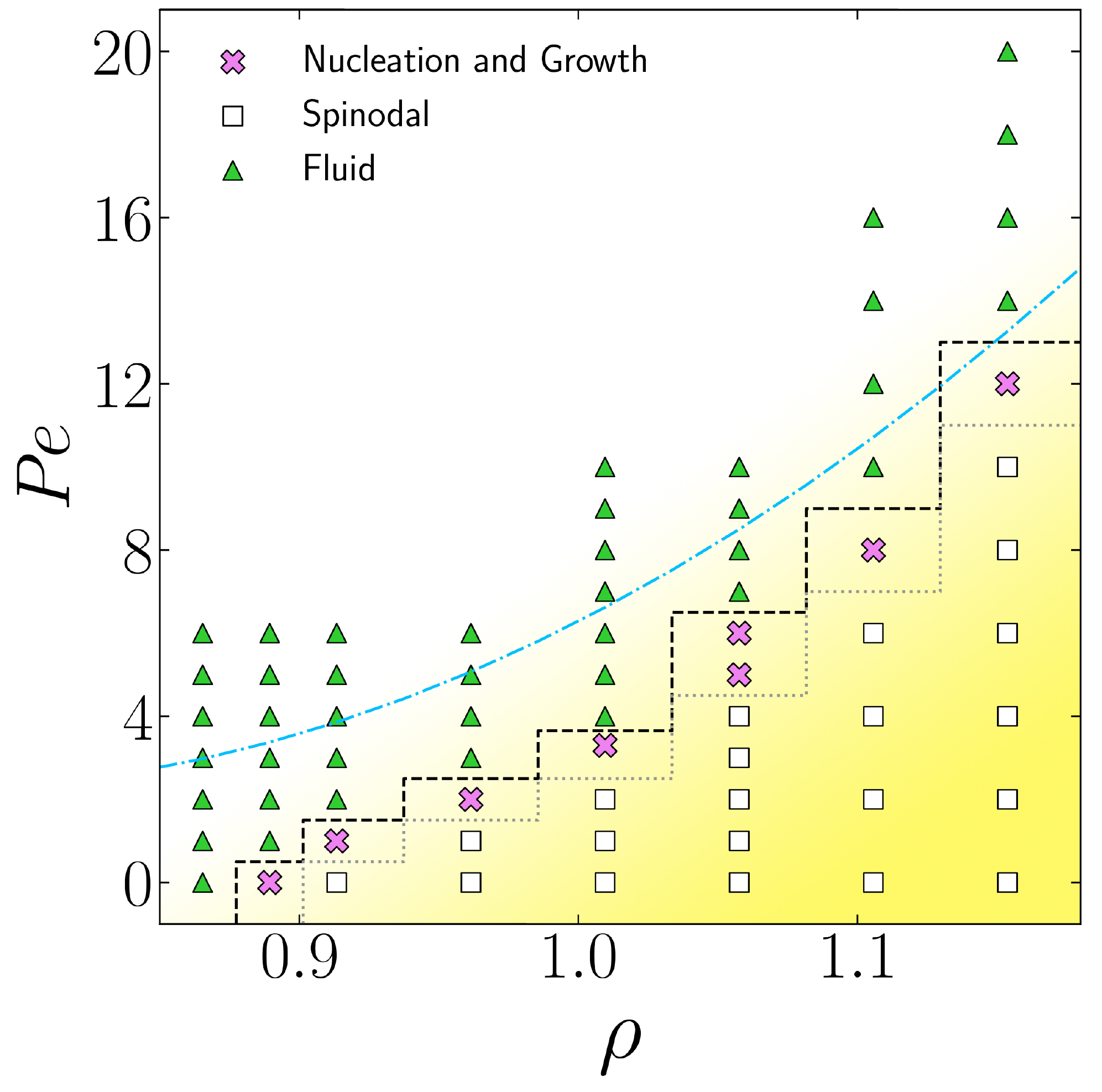}
\caption{State diagram showing crystallisation regimes of ABPs in 3D: Non-crystalline, ie. fluid states (green triangles), crystal freezing via nucleation and growth (pink crosses), and crystal freezing via spinodal growth (white squares). The freezing transition follows the dashed black line, spinodal and nucleation and growth regimes are separated by the dotted grey line. The blue dashed line marks states where the structural relaxation time is constant at $\tau_{\alpha}$ = 0.1. Crystalline states are defined as having cluster populations greater than 20\%.}
\label{figStateDiagram}
\end{figure}

\subsection{State Diagram}
\label{sectionStateDiagram}

In Fig. \ref{figStateDiagram}, we show the state points we consider and the results of our simulations. 
While other work has addressed the phase diagram of active Brownian particles in two and three dimensions with respect to MIPS \cite{richard2018,siebert2017}, or reported the full phase diagram of active disks, in which the freezing line is affected by activity \cite{digregorio2018};
here we distinguish the crystallisation regimes of nucleation and growth and spinodal by inspection of the crystallinity as a function of time data (see section \ref{sectionCrystalGrowthActivity}).
In particular, we identify behaviour compatible with the passive WCA system for $\pecl=0$. Recall that we carry out brute force simulations of $N\geq18000$ and for a run time of $7200\tau_R$. Therefore we do not obtain the equilibrium phase diagram \cite{kawasaki2010,filion2011}. Rather, for the passive case, we find nucleation and growth at density $\rho=0.89$ and spinodal crystallisation at $\rho=0.91$.
Moreover, since in the passive case we observe an equilibrium system, we can convert these densities to effective volume fractions $\phi$, via the Barker--Henderson method \cite{barker1967}. This comes out as $\phi=0.56$ for nucleation and growth and $\phi=0.57$ for spinodal crystallisation, which is consistent with previous work \cite{taffs2013,zaccarelli2009,valeriani2012,sanz2014,yanagishima2017}, noting that for numerical work such as this system size and runtime have significant consequences. We find similar trends to previous work which considers the effect of activity in two \cite{bialke2012,digregorio2018} and three dimensions \cite{wysocki2014,stenhammar2014,omar2021,turci2021} in which the freezing line moves to higher volume (or area) fraction as a function of activity.
We also note that the boundary between nucleation-and-growth and spinodal crystallization depends on system size, as large system sizes have a lower nucleation time shifting the transition between the two regimes to lower volume fractions (or generally to lower supercoolings~\cite{espinosa2016time}).

\begin{figure}
\includegraphics[width=0.8\linewidth]{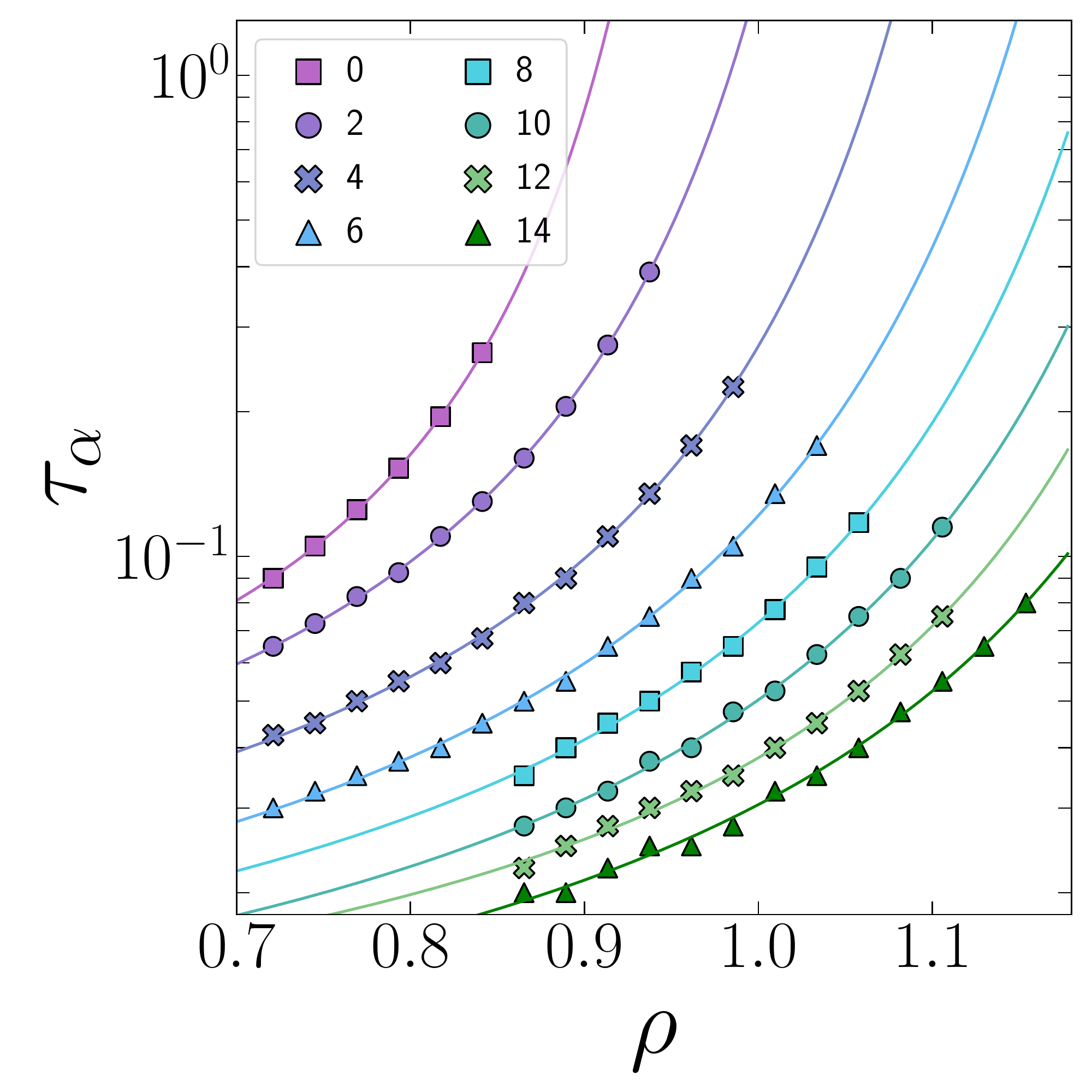}
\caption{Angell plot of structural relaxation time $\tau_\alpha$ as a function of $\rho$ , plotted for $\pecl$ 0 $\rightarrow$ 14 (see legend). Data collected from non-crystalline states.}
\label{figAngell}
\end{figure}

\subsection{Dynamical and Structural Response of the active WCA fluid to Activity}
\label{sectionDynamicalStructuralResponse}

\begin{figure*}
\includegraphics[width=140 mm]{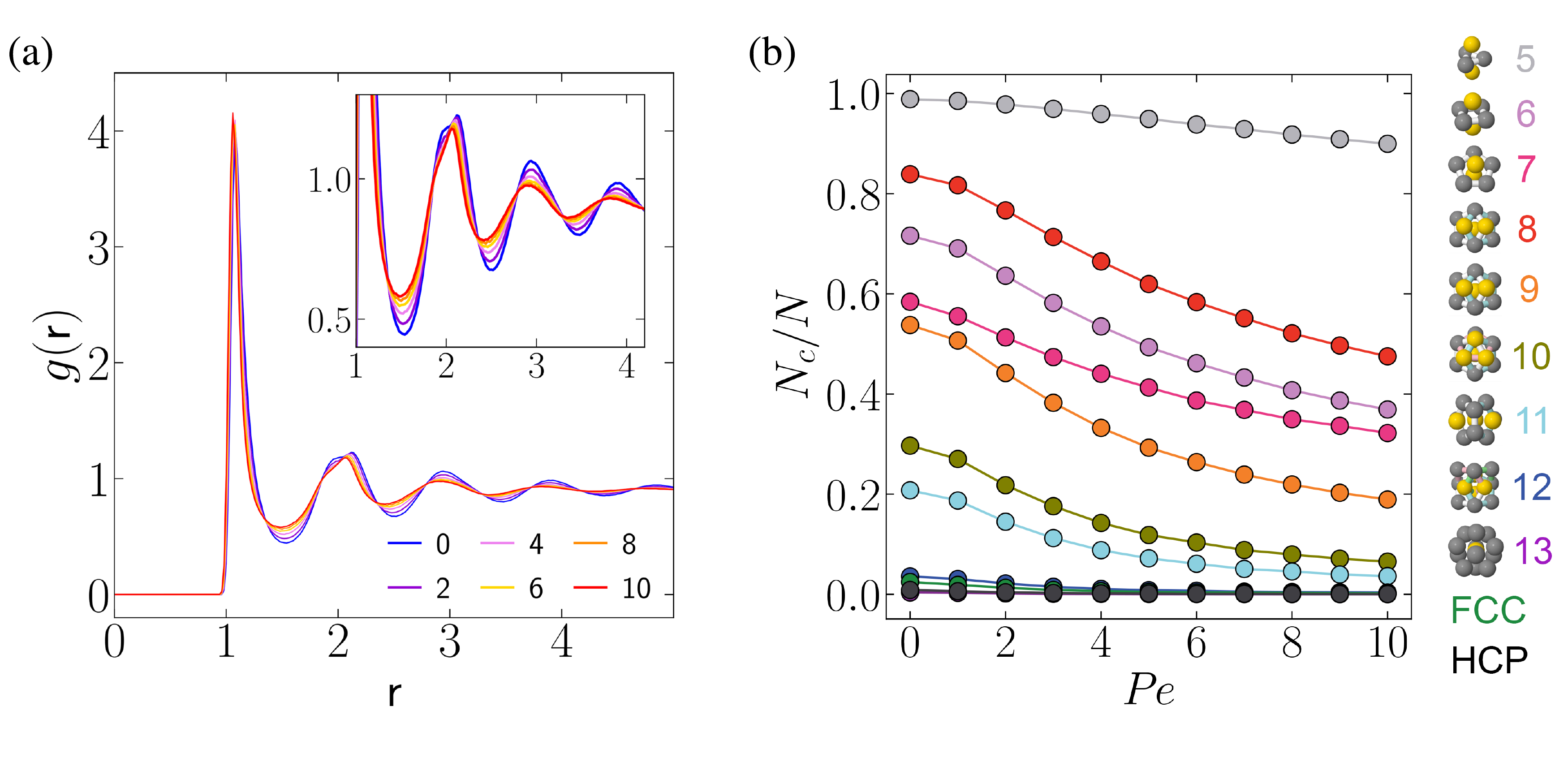}

\caption{Local structure in spherical ABPs at $\rho=0.87$. (a) Radial distribution function $g(r)$, plotted for $\pecl$=0, 2, 4, 6, 8, 10. (b) Cluster population as a function of $\pecl$. Colours correspond to the clusters depicted in the legend. With the increase of activity in the system, we observe increasingly less structure in the active fluid across all clusters. }
\label{figGR_TCC}
\end{figure*}

The intrinsic dynamics play an important role in setting the timescale of crystallisation. In this context the dynamical response of supercooled liquids to activity has been found to be highly complex and to exhibit qualitatively different responses to activity, from accelerating to slowing down and even non--monotonic behaviour \cite{janssen2019,szamel2015,berthier2017,dougan2016}. At the densities we consider ($\rho=0.72$ to $\rho=1.15$), in Fig. \ref{figAngell}, we see that upon increasing activity, the system accelerates and the structural relaxation time drops. Note that since we consider a monodisperse system, strong supercooling is not possible as crystallisation intervenes.

In Fig. \ref{figGR_TCC}(a) we show the two--body structure of the active fluids via the radial distribution function $g(r)$. The relationship between two--body structure and dynamics has been analysed in some detail in active systems \cite{janssen2019,szamel2015,berthier2017,dougan2016} and we see a familiar trend here, of a weakening of the strength of correlations as the relaxation time falls, which here is driven by an increase in activity (Fig. \ref{figAngell}).
This is particularly evident in the first minima and subsequent maxima and minima in the inset of Fig. \ref{figGR_TCC}(a).

We also consider the response of higher--order structure to activity in Fig. \ref{figGR_TCC}(b). Previously this has been found to develop as an increase in the population of locally favoured structures 
in the Wahnstr\"{o}m binary Lennard--Jones model, with activity induced via an Ornstein--Uhlenbeck process \cite{dougan2016}. However in Fig. \ref{figGR_TCC}(b), we find that the population of all local structures that we consider (those pertinent to the Lennard--Jones model \cite{wales1997,taffs2010}) \emph{decreases} with increasing activity for $\rho=0.87$ in which we do not find any crystallisation and simply focus on the liquid local structure. This decrease in higher--order structure is in marked contrast to the previous work with the Wahsntr\"{o}m model, however the latter, a model glassformer was much more deeply supercooled and the dynamics, like the higher--order structure exhibited the opposite response to activity noted here, suggesting these two systems are in different regimes according to the categorisation introduced in ref. \cite{berthier2017}. Our work shows qualitatively similar behaviour to passive systems when the temperature is increased \cite{royall2015physrep,taffs2010,malins2013fara}.


\begin{figure*}
\includegraphics[width=140 mm]{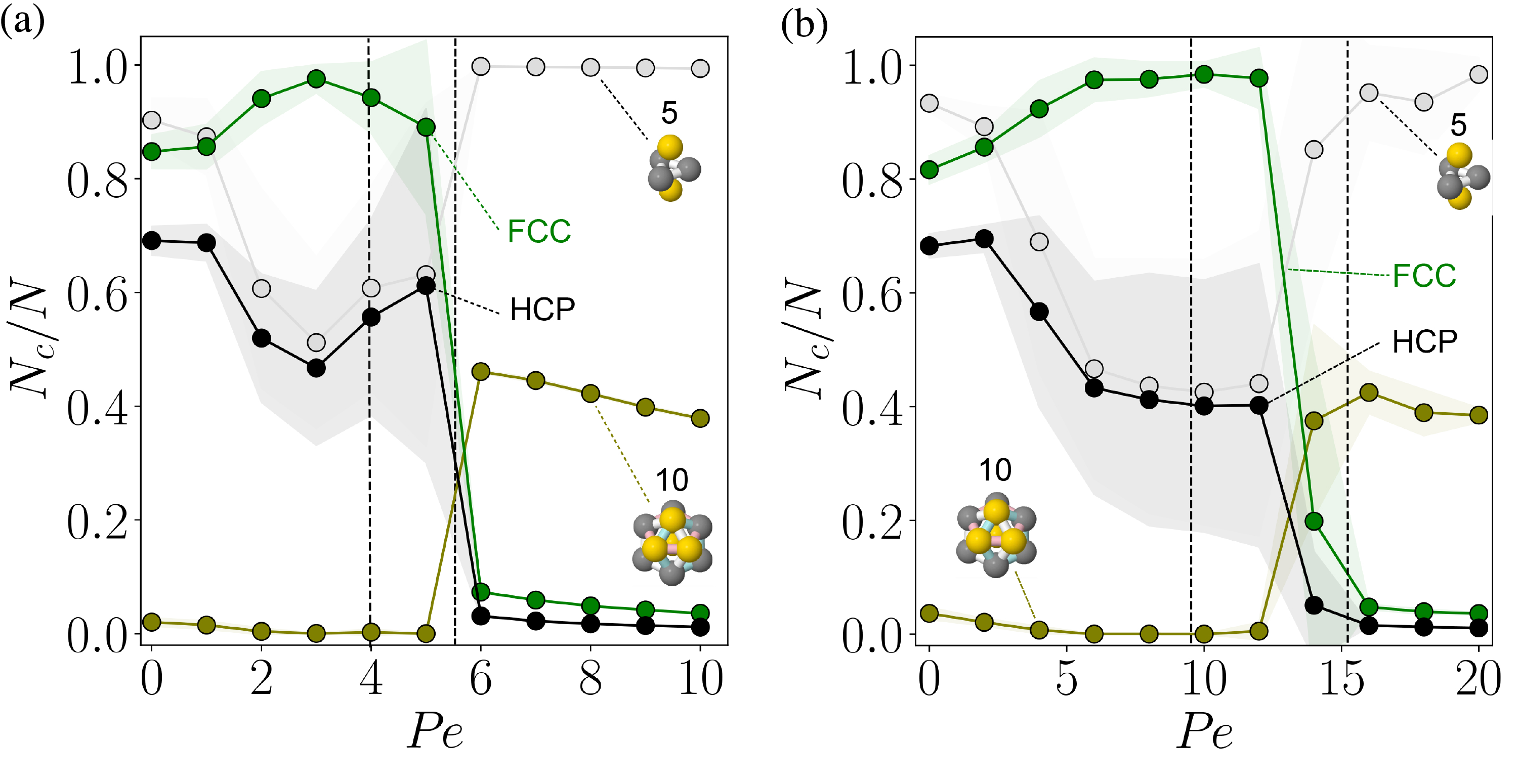}
\caption{Local structure in 3D ABPs
at two densities: $\rho=1.06$ (a), $\rho=1.15$ (b). Shaded regions show the standard deviation from 10 independent simulations, where the cluster populations averaged over configurations at $t=7200\tau_R$. Here we emphasise the 10--membered defective icosahedron among the amorphous local structures detected by the TCC because it is a locally favoured structure in the hard sphere system \cite{royall2015,hallett2018}. The dashed vertical lines signal the transition from spinodal growth to nucleation and growth, followed by the transition to the fluid regime as $\pecl$ increases.
\label{figPolymorph}
}
\end{figure*}

\subsection{Crystal Growth and Activity}
\label{sectionCrystalGrowthActivity}

We know from previous work \cite{wysocki2014,omar2021} that mono-disperse suspensions of active particles can crystallise at high density for activities below a certain Pe. The state diagram for this system is displayed in Fig.\ref{figStateDiagram}. We now consider in more detail the mechanism of crystal nucleation and polymorph selection at the examined state points.

In Fig.~\ref{figPolymorph} we look at the size of some selected cluster populations as a function of Peclet number for two densities in the region of stability of the solid phase, $\rho=1.06$ (a) and $\rho=1.15$ (b). In particular the five--membered triangular bipyramid consists of two tetrahedra (the simplex for spheres in 3d). We also consider the defective icosahedron which is a locally favoured structure of the hard sphere system \cite{royall2015,hallett2018}. All curves are obtained by averaging the final state of the simulation runs over 10 independent trajectories. We observe the following common trends with increasing activity ($\pecl$): For the passive case ($\pecl=0$) all trajectories crystallise into a mixture of FCC and HCP crystals, with a small preference for the FCC phase. This behaviour was observed in event driven simulations of hard-spheres and is explained by finite-size structural fluctuations that favour FCC-rich nuclei compared to HCP-rich nuclei, due to the higher stacking entropy of cubic phases compared to hexagonal phases.

Crystallisation at these high densities occurs spinodally, i.e. it is characterised by the appearance of multiple nucleation events, and where crystal growth is controlled by the annealing of stacking faults. Spinodal crystallisation persists when activity is introduced in the system. Looking at the FCC and HCP populations we observe that the effect of activity on polymorphism is to increase the fraction of FCC crystals with increasing $\pecl$, at the expense of the HCP population, which decreases with increasing $\pecl$. To explain the preference towards FCC we recall that the formation of hard-sphere crystals is subject to a mechanical instability under the effect of an external force which promotes the rearrangement of HCP layers into FCC layers \cite{heitkam2012}. This is confirmed in our simulations, where we observe the annealing of HCP stacking faults in favour of FCC environments promoted by the persistent motion of the active particles.
Interestingly the polymorph composition of the nuclei changes behaviour at a finite value of $\pecl$: for example, for $\rho=1.06$ ($\rho=1.15$) the FCC population reaches a maximum in relative composition at $\pecl\sim 3$ ($\text{Pe}\sim 8$).

This change of polymorphic behaviour coincides with a change in the crystallisation channel from spinodal to a nucleation-and-growth regime. In Fig.~\ref{figPolymorph} the onset of the nucleation-and-growth regime is indicated with the dashed vertical line. Here nucleation is a rare event and within our simulation box the crystal grows from a single critical nucleus. In this regime, nucleation proceeds with a smaller number of grain boundaries, and the polymorph composition tends towards the passive value.

Further increasing the activity causes the nucleation rate to drop, until crystallisation is no longer observed. In Fig.~\ref{figPolymorph} this transition is represented with the vertical dash-dotted line where, not only do crystalline environments rapidly decay, but defective icosahedra environments increase to signify the transition to a fluid regime. At high $\pecl$, this higher--order structure weakens, similar to the fluid case ($\rho=0.87$) [Fig. \ref{figGR_TCC}(b)].

\begin{figure}
	\includegraphics[width=0.8\linewidth]{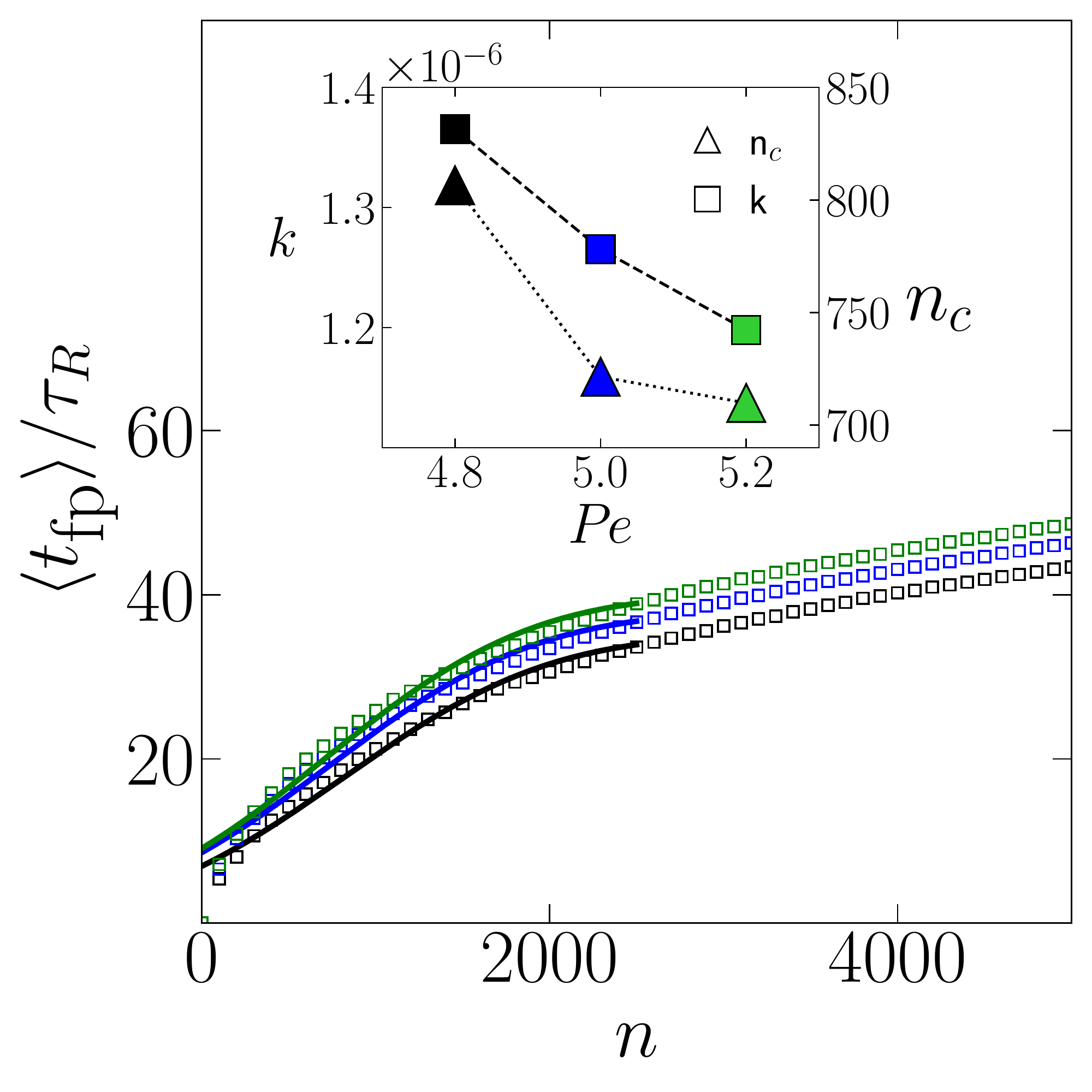}
	\caption{Mean first passage time as function of nucleus size $n$ in the nucleation and growth regime: $\rho=1.06$, Pe = 4.8, 5.0, 5.2 plotted in black, blue and green respectively. Inset shows the nucleation rate $k$ and the critical nucleus $n_c$ as a function of $\pecl$, extracted from fitting at low $n$.}
		\label{figMFPT}
\end{figure}

\begin{figure*}
\includegraphics[width=0.8\linewidth]{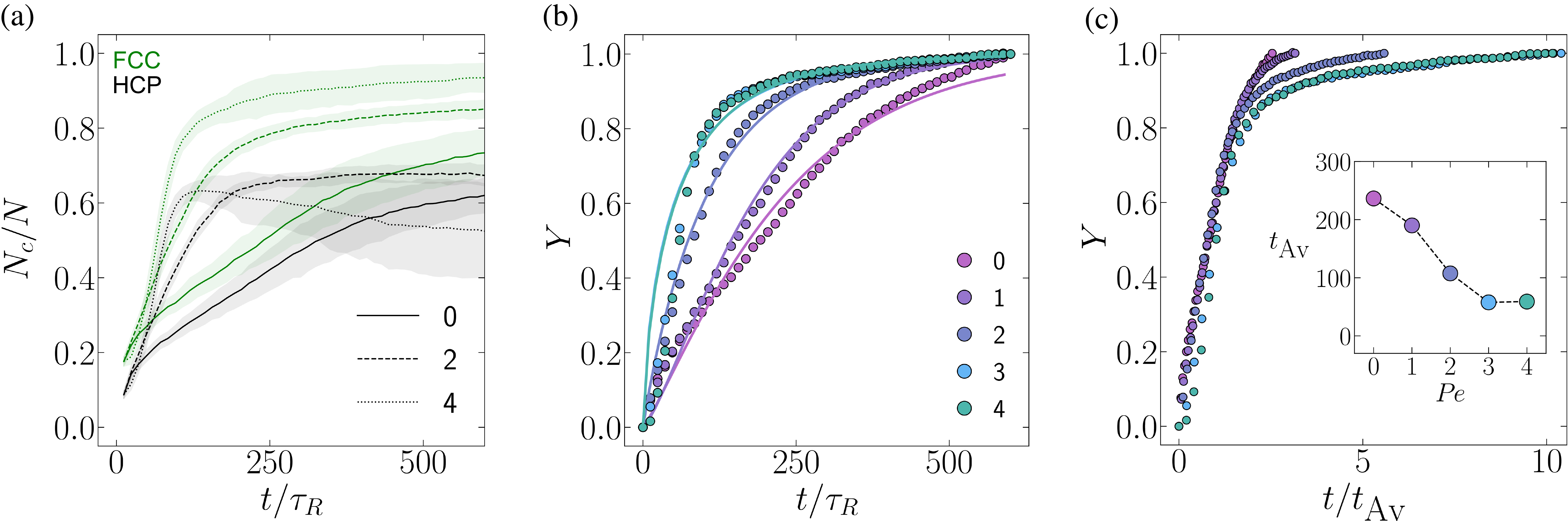}

	\caption{(a) FCC and HCP cluster population growth time in the spinodal regime for $\rho=1.06$, and $\pecl =0,2,4$. FCC and HCP are plotted in green and black respectively and the $\pecl$ is denoted by the line-style. Shading indicates the standard deviation from averaging 20 independent simulations. (b) FCC crystal fraction $Y$ for $\pecl =0,1,2,3,4$ at $\phi=0.67$, and  here $Y$ has been fit with the Avrami eqaution. (c) as for (b) but with $Y$ scaled by the characteristic time $t_{Av}$; inset shows the variation of $t_{Av}$ with $\pecl$.}
		\label{figMeanPopVsT}
\end{figure*}

In Fig.~\ref{figMFPT} we focus on the state points displaying nucleation-and-growth and plot the mean first passage time $\langle t_\text{fp}(n)\rangle$, defined as the average elapsed time until the appearance of a nucleus of size $n$, 
at $\rho=1.06$ and at three different $\pecl$ numbers. Over a wide range of $n$, $\langle t_\text{fp}(n)\rangle$ can be fitted with the expression

\begin{equation}\label{eqn:mfpt}
\langle t_\text{fp}(n)\rangle = \frac{1}{2 k V} \{1+\text{erf} \left[c(n-n_c)\right]\}
\end{equation}
where $k$ is the nucleation rate, $n_c$ is the critical nucleus size, $\text{erf}$ is the error function, and $c$ is a constant which in the equilibrium case ($\pecl=0$) is proportional to the curvature of the nucleation barrier $\Delta F$ at the critical size, $c=\sqrt{\Delta F''(n_c)/k_BT}$.
The curves show that the mean first passage time increases with increasing activity, with a consequent drop in the nucleation rates $K$ extracted from the functional fits of Eq.~\ref{eqn:mfpt} and plotted in the inset. Activity hinders nucleation, and even more so if the nucleation rates are scaled by the relaxation time in the fluid $\tau_\alpha$ which, as plotted in Fig.~\ref{figAngell}, drops faster than exponentially with increasing $\pecl$. Interestingly the critical nucleus size $n_c$, as indicated on the right axis 
of Fig.~\ref{figMFPT} inset, also decreases with increasing activity. These critical sizes are considerably larger than what is typically observed in the passive case (at the same density), owing to the acceleration of the underlying fluid dynamics with activity, that allows the observation of longer nucleation induction times.

In Fig.~\ref{figMeanPopVsT} we focus on the spinodal nucleation regime, i.e. when the non-equilibrium nucleation barrier is low enough for multiple nucleation events to occur simultaneously in the simulation box, and crystallisation is an activated process controlled by the rate of addition of new crystals on the nuclei. Panel (a) shows the fraction of crystalline particles as a function of time for $\pecl=0,2,4$ (continuous, dashed and dotted curves respectively) and distinguishing between FCC (green curves) and HCP (black curves). For both FCC and HCP we observe an increase of the crystal growth rate as a function of activity. The fraction of HCP at $\pecl=4$ shows a marked decrease at long times, which is due to the annealing of HCP stacking faults that we observed also in Fig.~\ref{figPolymorph}. To analyse the growth regime in panel (b) we fit the curves for the FCC phase with the Avrami equation.

\begin{equation}
Y=1-e^{-K t^n}
\end{equation}
where $Y$ is the crystal fraction $Y=(N_c-N_0)/(N-N_0)$, with $N_c$ the number of crystalline particles, $N_0$ the starting number of crystalline particles, and $N$ the total number of particles. $K=\pi k\dot{G}^3/3$ is the Avrami constant proportional to the nucleation rate $k$ and the growth rate $\dot{G}$, and $n$ is the Avrami exponent. From the fits in panel (b) we obtain $n\simeq 1$ and an increase in the growth constant $K$ with increasing $\pecl$. In panel (c) we show how the crystalline growth is rescaled by a characteristic time $t_\text{Av}=K^{-1/n}$ (plotted in the inset). This timescale decreases with activity, signalling the increase in the growth rate with $\pecl$. What appears to be an increase in the growth rate of nuclei in units of the active particles rotational time $\tau_R$, is still a significant slowing down if measured instead in units of the relation time $\tau_\alpha$.

\label{sectionCrystalGrowthActivity}

\section{Conclusion}
\label{sectionConclusion}

We have considered the crystallisation behaviour of a suspension of active Brownian Particles that interact with a hard-sphere like interaction. We showed that the freezing line is strongly affected by activity, and moves to higher densities as we increase the $\pecl$ number of the active particles consistent with previous work \cite{wysocki2014,stenhammar2014}. This is accompanied by a reduction of the nucleation rate, nucleation barriers and critical nucleus size with increasing activity. Despite the suppression of nucleation, the growth of nuclei is enhanced by the accelerated dynamics of the melt. This allows us to observe spinodal nucleation, where the growth is controlled by the rate of particle attachment, and thus speeded-up with activity.

We observe a decrease in pair-- and higher--order structure in the fluid with increasing activity. The former is compatible with certain dynamic regimes observed previously \cite{janssen2019,szamel2015,berthier2017}. This is intriguing as one may enquire as to the nature of the higher--order structure approaching the MIPS phase boundary \cite{omar2021,turci2021}. Very recently, comparisons have been made between MIPS and criticality in passive systems \cite{turci2021}, and in the case of passive systems, approaching criticality, the population of higher--order structure detected by the TCC \emph{increases}  \cite{richard2018}, in marked contrast to our findings here. In the future, it would be interesting to investigate whether the trend we have observed changes closer to the MIPS boundary or whether the response of he higher--order structure is profoundly different to passive systems.

Remarkably activity also has a strong effect on polymorph selection. While the passive system crystallises in an equimolar mixture of FCC and HCP, active particles progressively favour the FCC phase at higher $\pecl$. We observe this as annealing of HCP stacking faults, especially close to the crystal boundaries.

\begin{acknowledgments}
We thank Silke Henkes and Francesco Turci for many valuable discussions. F.J.M. is supported by a studentship provided by theBristol Centre for Functional Nanomaterials (EPSRCgrant EP/L016648/1). 
J.R. acknowledges support from the European Research Council Grant DLV-759187.
C.P.R. gratefully acknowledges the Royal Society, European Research Council (ERC Consolidator Grant NANOPRS, project number 617266) and EPSRC EP/T031077/1.
\end{acknowledgments}

\section*{Author Contributions}
All simulations and numerical analysis were performed by FJM. All authors wrote the manuscript and analysed data.


%

\end{document}